\documentclass[epjST]{svjour}
\usepackage{graphics}
\usepackage{epsfig}
\usepackage{graphicx}
\usepackage{epsfig,color}
\usepackage{cite}
\usepackage[english]{babel}
\usepackage{amsmath}
\usepackage{floatrow}
\usepackage[label font=bf]{subfig}
\usepackage{caption}
\usepackage{siunitx}
\usepackage{xcolor}
\usepackage{mathabx}
\usepackage{verbatim}

\begin{document}
\title{Firing rate model for brain rhythms controlled by astrocytes}
\author{Sergey V. Stasenko\inst{1}\fnmsep \and Sergey M. Olenin\inst{1}\fnmsep \and  Eugene A. Grines\inst{1}\fnmsep \and Tatiana A. Levanova \inst{1}\fnmsep\thanks{\email{tatiana.levanova@itmm.unn.ru}}  }
\institute{Lobachevsky State University of Nizhny Novgorod, 23 Prospekt Gagarina, Nizhny Novgorod, 603950, Russia}
\abstract{
We propose a new mean-field model of brain rhythms governed by astrocytes. This theoretical framework describes how astrocytes can regulate neuronal activity and contribute to the generation of brain rhythms. The model describes at the population level the interactions between two large groups of excitatory and inhibitory neurons. The excitatory population is governed by astrocytes via a so-called tripartite synapse. This approach allows us to describe how the interactions between different groups of neurons and astrocytes can give rise to various patterns of synchronized activity and transitions between them. Using methods of nonlinear analysis we show that astrocytic modulation can lead to a change in the period and amplitude of oscillations in the populations of neurons.
} %end of abstract
\maketitle
\section{Introduction}
\label{intro}

Brain rhythms, also known as neural oscillations or brain waves, are the patterns of electrical activity in the brain. They are believed to play a crucial role in the information processing within the brain, and have been linked to a variety of cognitive and behavioral functions \cite{buzsaki2004neuronal}. Different types of brain rhythms are associated with different cognitive processes. For example, the theta rhythm (4-7 Hz) has been linked to working memory and spatial navigation, while the gamma rhythm (30-80 Hz) has been associated with attention and perception \cite{jensen2010shaping}. There are several different ways in which brain rhythms can be regulated: neuronal feedback \cite{buzsaki2012mechanisms}, neuromodulation \cite{hasselmo2014theta}, environmental factors \cite{buzsaki2022brain} and control by glial cells  \cite{poskanzer2016astrocytes}. The last factor is the least studied one.

The generation and regulation of brain rhythms is a complex process that involves the interactions between many different types of cells in the brain, including neurons and glial cells. Glial cells, which include astrocytes, oligodendrocytes, and microglia, were previously thought to be merely supportive cells for neurons. However, recent research has revealed that glial cells play a much more active role in the brain functioning than it was previously thought \cite{verkhratsky2018physiology}. Astrocytes make the greatest contribution to the regulation of synaptic transmission among all glial cells \cite{Araque1998,Araque1999,Wittenberg2002,Wang1999}.  An emerging study proposed that astrocytes could potentially impact the characteristics of cortical gamma oscillations \cite{lee2014astrocytes,makovkin2022controlling,vodovozov2018metabolic,perea2016activity,tan2017glia}. Specifically, astrocytes play crucial role in sustaining of functional gamma oscillations and in novel object recognition behavior, both in controlled laboratory settings and in awake animals \cite{lee2014astrocytes}. 

It is known that astrocytes form bidirectional connections with neurons, using which it can affect the pre- and postsynaptic compartments of the synapse by releasing gliotransmitters in a calcium-dependent manner \cite{Wittenberg2002,Araque1999,Haydon2001}. This framework is known as a tripartite synapse. When part of the neurotransmitter released from the presynaptic terminal binds to corresponding receptors on the astrocyte membrane, a cascade of biochemical reactions occurs \cite{Perea2009}. As a result, gliotransmitters are released into the synaptic cleft and extrasynaptic space. Described process allows to modulate synaptic transmission. Astrocytes have been proposed to act as frequency-selective "gate keepers" and presynaptic regulators \cite{Nadkarni2004,Nadkarni2007}. Corresponding gliotransmitters can effectively modulate presynaptic facilitation and depression \cite{volman2007astrocyte,DePitta2011}. 

The functional role of astrocytes in neuronal dynamics has been extensively studied using different types of mathematical models \cite{Postnov2007,Amiri2011,Wade2011,amiri2013astrocyte,pankratova2019neuronal}. Numerous studies on network and brain modeling approaches employ mean-field models. Their relative simplicity allows studying network dynamics at macroscopic level. On the example of mean-field model of tripartite synapse in a number of studies (see, e.g., \cite{stasenko2023loss,stasenko2023dynamic,stasenko2023information,zimin2023artificial,gordleeva2021modeling}) it has been investigated how astrocytes participate in the coordination of neuronal signaling, particularly in spike-timing-dependent plasticity and learning. Extensive studies \cite{Gordleeva2012,de2019gliotransmitter,
lazarevich2017synaptic,stasenko2020quasi,barabash2021stsp,
stasenko20223d,barabash2023rhy,olenin2023dynamics} on astrocytic modulation of neuronal activity reveal that functional gliotransmission is a complex phenomenon that depends on the nature of structural and functional coupling between astrocytic and synaptic elements. Nevertheless, the precise mechanisms through which astrocytes contribute to the synchronized network activity and brain rhythms are far from being fully understood \cite{lenk2020computational}.

In order to shed a light into some of these mechanisms we propose a new mean-field model that describes the collective behavior of two large population of neurons in the presence of astrocytes. The proposed model phenomenologically describes how astrocytes modulate the excitability of neurons through the release of gliotransmitters, which can either enhance or inhibit neuronal activity. We show that such modulation can lead to a change in the period and amplitude of oscillations in the population activity of neurons.

The paper is organized as follows. In Section \ref{sec:model} we introduce a mean-field model of two large neuronal populations, one of which is controlled by astrocytes. In Section \ref{sec:res} using analytical and numerical methods of nonlinear analysis and bifurcation theory we describe how astrocytes can regulate brain rhythms. In Section \ref{sec:conclusion} we discuss our findings and draw our conclusions.

\section{Two-population rate model with synaptic dynamics and astrocytic influence}
\label{sec:model}

Mean field theory provides a useful tool for analysis of collective behavior of a large populations of interacting units, allowing corresponding dynamics to be reduced to the properties of several parameters. In neural circuits, these parameters are typically the firing rates of distinct homogeneous populations of neurons. Knowledge of the firing rates under conditions of interest can reveal essential information on both the dynamics of neural circuits and the way they can subserve brain function \cite{la2021mean}. 

\begin{figure}[tbh]
    \centering
    \includegraphics[scale=0.7]{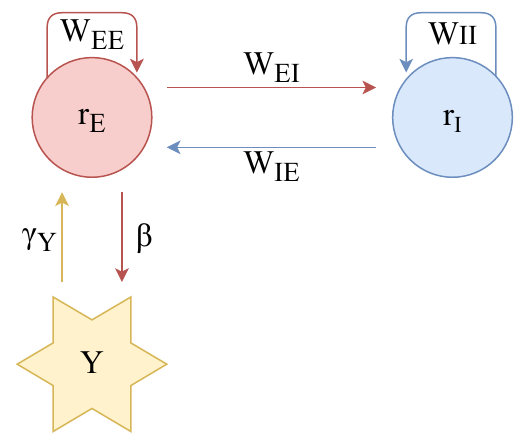}
    \caption{Overall scheme representing the phenomenon of interaction of excitatory and an inhibitory populations. Here $r_E(t)$ and $r_I(t)$ are firing rates for E and I populations, correspondingly. Weight matrices $w_{ab}$, where $a\in [E,I]$, $b\in [E,I]$ set weights of couplings inside one populations and between different populations. }
    \label{fig:scheme}
\end{figure}

Based on the results of experimental studies described in literature, e.g., in \cite{lebedeva2023effect}, we developed a new phenomenlogical mean-field model, which consist of two large populations of neurons: an excitatory (E) and an inhibitory (I) ones. The proposed model allows one to reproduce brain rhythms which emerge as a result of interplay between E and I populations under astrocytic influence. The model implements crucial features of neuronal populations in the brain, which allow to reproduce E-I-based gamma oscillations \cite{buzsaki2012mechanisms}: reciprocal connections between the excitatory and inhibitory populations, E-E self-connectivity within the excitatory population, and external stimuli applied to each population ($I_E$ and $I_I$). The schematic representation of the biological mechanism underlie E-I-based gamma oscillations is presented in Fig.~\ref{fig:scheme}.

The proposed model is based on the two-population rate model with synaptic dynamics \cite{keeley2019firing}. In its turn, it is based on the Wilson-Cowan formalism \cite{wilson1972excitatory}, which can be described by a 2-dimensional system of ordinary differential equations (ODEs). The distinguishing feature of our model is an additional description of the dynamics of neurotransmitters for excitatory (glutamate) and inhibitory (GABAergic) synapses, which are set as constant values in the original Wilson-Cowan model. We also introduce the dynamics of gliotransmitter (glutamate), which regulates the fraction of the released neurotransmitter in excitatory synapses. 

Mathematically the proposed rate model with astrocyte can be described using the following 5-dimensional system of ODEs:
\begin{equation}
\begin{array}{ll}
\tau_{s_{E}} \frac{ds_{E}}{dt}=-s_{E}+g(Y) \gamma_{E} r_{E}(1-s_{E})+s_{0}^{E}, \\
\\
\tau_{s_{I}} \frac{ds_{I}}{dt}=-s_{I}+\gamma_{I} r_{I}(1-s_{I})+s_{0}^{I}, \\\\
\tau_{r_{E}} \frac{dr_{E}}{dt}=-r_{E}+f_{E}(I_E + \omega_{EE}s_{E} - \omega_{IE}s_{I}), \\
\\
\tau_{r_{I}} \frac{dr_{I}}{dt}=-r_{I}+f_{I}(I_I + \omega_{EI}s_{E} - \omega_{II}s_{I}), \\ \\
\frac{dY}{dt} = \frac{-Y}{\tau_{Y}}  + \beta H_{Y}(s_E).\\
\end{array}
\label{eq:main}
\end{equation}
Here each population (excitatory or inhibitory) is characterized by a firing rate variable $r_a(t)$ and a synaptic variable $s_a(t)$, $a=[E,I]$. The synaptic activity depends on the population firing rate, which is scaled by $\gamma_a$ and saturates at 1, with a decay time constant $\tau_{s_a}$ and a background drive $s_0^a$. The synaptic rise time constant is $\tau_{s_a}/\gamma_a$. The rate is a function of synaptic weights $w_{ab}$ and a sigmoidal function $f$ of the following form: 
\begin{equation}
    f_{i}(arg) = \frac{1}{1 + \exp{(-k_{i}(arg-\theta_{i}))}}, \\
    i \in [E,I]
\end{equation}
with threshold constants $\theta_i$, $i \in [E,I]$. Note that self- and cross-population targets have no difference in synaptic activation dynamics, so $s_I$ is similar for inhibitory inputs to excitatory and inhibitory cells, and only the weights differ \cite{brunel2003determines,hansel2003asynchronous,kopell2010gamma}. 

The variable $Y(t)$ describes the dynamics of gliotransmitter release, with relaxation time $\tau_{Y}=1$ s and an activation function $H_{Y}(s_E)$:
\begin{equation}
H_{Y}(s_E) = \frac{1}{1 + e^{-(s_E-s_E{_{thr})/k_Y}}},
\end{equation} 
Astrocyte activity leads to the release of gliotransmitter, which binds to presynaptic neuron receptors and changes the neurotransmitter release. The model introduces astrocytic influence on glutamate release through the function $g(Y)$, which includes the coefficient of astrocyte influence on synaptic connection $\gamma_Y$ and an activation threshold $Y_{thr}$:
\begin{equation}
   g(Y) = 1  + \frac{\gamma_Y}{1 + e^{-Y+Y_{thr})}}.
\end{equation} 

In our study the parameters $\gamma_Y$, $I_E$, $I_I$, $s^E_0$, $s^I_0$ were chosen as the control parameter. The values of the remaining parameters were fixed as follows: %$s_0^E = 0.15$, $s_0^I = 0.1$, $I_E = 0.9$, $I_I = 0$,
$\tau_r^E = 2 \ \mathrm{ms}$, $\tau_r^I = 6 \ \mathrm{ms}$, $\tau_s^E = 3 \ \mathrm{ms}$, $\tau_s^I = 10 \ \mathrm{ms}$,  $\gamma_E = 4$, $\gamma_I = 8$, $w_{EE} = 3.5$, $w_{EI} = 5$, $w_{II} = 3$, $w_{IE} = 5$, $\gamma_Y = 0.305$, $\beta_Y = 10 \ \mathrm{ms^{-1}}$, $\tau_Y=10 \ \mathrm{ms}$, $\theta_E=0.2$, $\theta_I=0.4$. The parameters of the population neuron model are chosen in such a way that, without astrocytic influence, the system \ref{eq:main} demonstrate gamma oscillations (which corresponds to an oscillation frequency above $30 \mathrm{Hz}$) \cite{keeley2019firing}. Parameters related to astrocyte were chosen as it was done in several papers devoted to mean-fleld modelling of neuron-astrocyte interactions \cite{lazarevich2017synaptic,barabash2023rhy,olenin2023dynamics}. Most of the parameters in the model are dimensionless, as in the Wilson-Cowan model, with the exception of time constants.

\section{The results}
\label{sec:res}

This section presents the results of analytical and numerical study of the proposed model using methods of nonlinear analysis and bifurcation theory. In Subsection \ref{subsec:eq_states} we perform equilibrium states analysis. In Subsection \ref{subsec:bifurc} we conduct numerical two-parameter bifurcation analysis paying special attention to bifurcations that lead to formation and destruction of brain rhythms. In Subsection \ref{subsec:rhytms} we show examples of different types of rhythmic activity and discuss peculiarities of transformation of one rhythms to another.

\subsection{Analytical study}
\label{subsec:eq_states}

Here we will describe certain properties of equations \eqref{eq:main}. 
The most important property is that all biologically relevant (i.e., the ones that are entirely located in positive orthant $s_E, \, s_I , \, r_E, \, r_I, \, Y \geqslant 0$ and, hence, compatible with the biological meaning of the variables) equilibria and attractors of this system are always contained in a hyper rectangle $\mathcal{H}$ defined by inequalities $s_E, s_I, r_E, r_I \in \lbrack0, \, 1 \rbrack, \, Y \in \lbrack 0, \, \beta \cdot \tau_Y \rbrack$ provided that all parameter values are positive and $s_0^E, s_0^I < 1$. 
Indeed, let us check the direction of vector field at the following hyperplanes:
$$\left . \frac{ds_{E}}{dt} \right \vert _{s_E = v} = \frac{\gamma_{E} r_{E} \left(1 - v\right) g{\left(Y \right)} + s^{E}_{0} - v}{\tau_{s}^{E}},$$
$$\left . \frac{ds_{I}}{dt} \right \vert _{s_I = v} = \frac{\gamma_{I} r_{I} \left(1 - v\right) + s^{I}_{0} - v}{\tau_{s}^{I}},$$
$$\left . \frac{dr_{E}}{dt} \right \vert _{r_E = v} = \frac{- v + f_{E}{\left(I_{E} + s_{E} w_{EE} - s_{I} w_{IE} \right)}}{\tau_{r}^{E}},$$
$$\left . \frac{dr_{I}}{dt} \right \vert _{r_I = v} = \frac{- v + f_{I}{\left(I_{I} + s_{E} w_{EI} - s_{I} w_{II} \right)}}{\tau_{r}^{I}}.$$
Clearly, all of these derivatives are positive, when $v = 0$, due to positivity of parameters, functions $f_E$ and $f_I$ having a range of $(0, 1)$ and due to considering only non-negative values of phase state variables (non-negativity of $r_E$ and $r_I$ is crucial for derivatives $\left . \frac{ds_{E}}{dt} \right \vert _{s_E = 0}$ and $\left . \frac{ds_{I}}{dt} \right \vert _{s_I = 0}$).
Thus, the direction of the vector field on the boundary of positive orthant forces trajectories to stay in it.
For similar reasons all of these four derivatives are negative, when $v \geqslant 1$. 
As a consequence, if a phase trajectory starts at the point with, for example, $s_E > 1$, the value of this phase variable will decrease until it gets into unit segment $\lbrack0, \, 1\rbrack$.
Also, if a trajectory starts with values of phase variables $s_E, s_I, r_E, r_I \in \lbrack0, \, 1 \rbrack$, the vector field direction at the boundary forbids trajectories to get out of these limits. 
Similar conclusions hold for the derivative of $Y$ variable:
$$\left . \frac{dY}{dt} \right \vert _{Y = v}= \beta H_{Y}{\left(s_{E} \right)} - \frac{v}{\tau_{Y}},$$
which is positive, when $v = 0$, and negative, when $v \geqslant \beta \cdot \tau_Y$. 

Combining all these properties we arrive at the conclusion that even trajectories that start in the positive orthant, but outside of the hyper-rectangle $\mathcal{H}$, eventually reach it in forward time, thus all attractors must be contained in $\mathcal{H}$.
The same derivative estimates prevent an occurrence of any equilibria states in positive orthant outside of hyper-rectangle $\mathcal{H}$.
These properties allow us to sample only a compact portion of phase space when searching for equilibria or attractors, which is presented further in this paper. 

Let us also describe a special structure of system \eqref{eq:main} that allows to simplify a search for its equilibria. 
From the equation $\frac{dY}{dt} = 0$ we can express $Y$ as a function of $s_E$ 
$$ Y = \beta \tau_{Y} H_{Y}{\left(s_{E} \right)} $$
and plug that into the equation $\frac{ds_{E}}{dt} = 0$. 
After that we can express $r_E$ as a function of $s_E$
$$ r_E = \frac{- s^{E}_{0} + s_{E}}{\gamma_{E} \left(1 - s_{E}\right) g{\left(\beta \tau_{Y} H_{Y}{\left(s_{E} \right)} \right)}}, $$
while from $\frac{ds_{I}}{dt} = 0$ we will express $r_I$ as a function of $s_I$
$$r_I = \frac{- s^{I}_{0} + s_{I}}{\gamma_{I} \left(1 - s_{I}\right)} .$$
Finally, plugging these expressions into the equations $\frac{dr_{E}}{dt} = 0$ and $\frac{dr_{I}}{dt} = 0$ gives a reduced system of equations involving only variables $s_E$ and $s_I$:
\begin{equation}
\label{eq:TruncEquilibria}
\left \lbrace 
\begin{matrix}
\gamma_{E} \left(1 - s_{E}\right) f_{E}{\left(I_{E} + s_{E} w_{EE} - s_{I} w_{IE} \right)} g{\left(\beta \tau_{Y} H_{Y}{\left(s_{E} \right)} \right)} + s^{E}_{0} - s_{E} = 0,\\
\gamma_{I} \left(1 - s_{I}\right) f_{I}{\left(I_{I} + s_{E} w_{EI} - s_{I} w_{II} \right)} + s^{I}_{0} - s_{I} = 0.
\end{matrix}
\right .
\end{equation}
As we know from the previous paragraph, neither of the derivatives vanish outside of hyper-rectangle $\mathcal{H}$, hence we can restrict $s_E$ and $s_I$ to the unit segment $\lbrack 0, \, 1 \rbrack$. 
A solution $(s_E^\ast, \, s_I^\ast)$ of the system \eqref{eq:TruncEquilibria} can be used to find other coordinates of the equilibrium state:
\begin{equation}
\label{eq:BackSubst}
\begin{array}{rcl}
Y & = & \beta \tau_{Y} H_{Y}{\left(s_{E}^\ast \right)},\\
r_E & = & \frac{- s^{E}_{0} + s_{E}^\ast}{\gamma_{E} \left(1 - s_{E}^\ast\right) g{\left(\beta \tau_{Y} H_{Y}{\left(s_{E}^\ast \right)} \right)}}, \\
r_I & = & \frac{- s^{I}_{0} + s_{I}^\ast}{\gamma_{I} \left(1 - s_{I}^\ast\right)} .
\end{array}
\end{equation}
The formulas \eqref{eq:BackSubst} suggest even a tighter search region for the solutions of \eqref{eq:TruncEquilibria}: if $s_E^\ast < s_0^E$ or $s_I^\ast < s_0^I$, they will give rise to a non-biological equilibrium state with negative values of either $r_E$ or $r_I$. 

While occurrence of such equilibria does not contradict previous observations (such equilibria are located outside of positive orthant and earlier the derivative estimations  were given under an assumption of non-negativity of phase variables), they still have to be excluded from the consideration due to being in a part of phase space that has no biological meaning.

Thus, the search region for the solution of equations \eqref{eq:TruncEquilibria} can be narrowed to $s_E \in \lbrack s_0^E, \, 1 \rbrack$, $s_I \in \lbrack s_0^I, \, 1 \rbrack$.

\subsection{Two-parameter numerical bifurcation analysis}
\label{subsec:bifurc}

\begin{figure}
\begin{center}
    (a)\includegraphics[width = 0.45\linewidth]{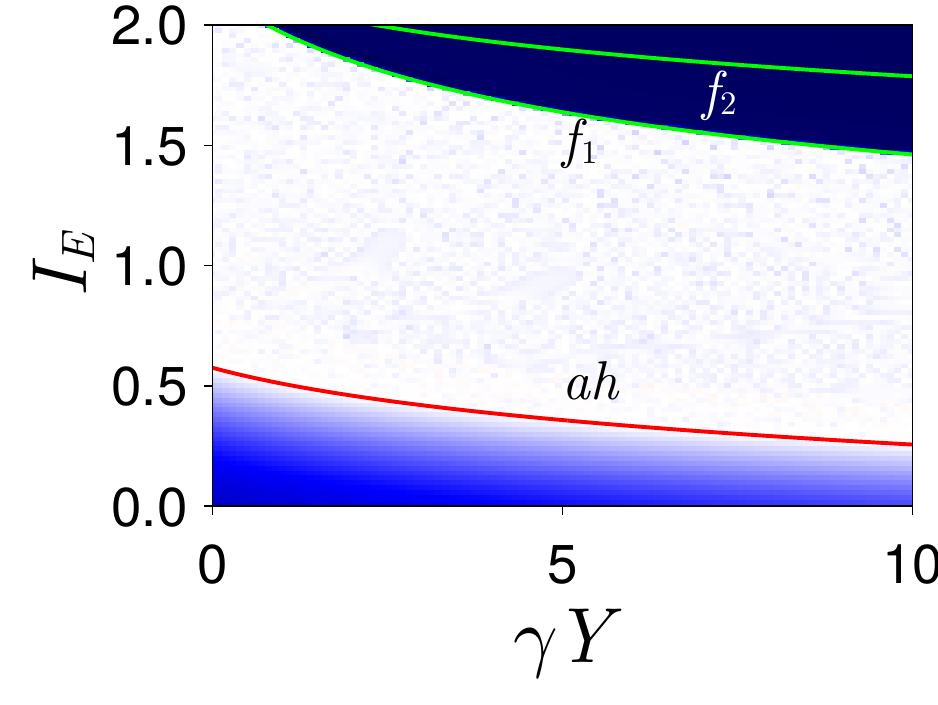}
    (b)\includegraphics[width = 0.45\linewidth]{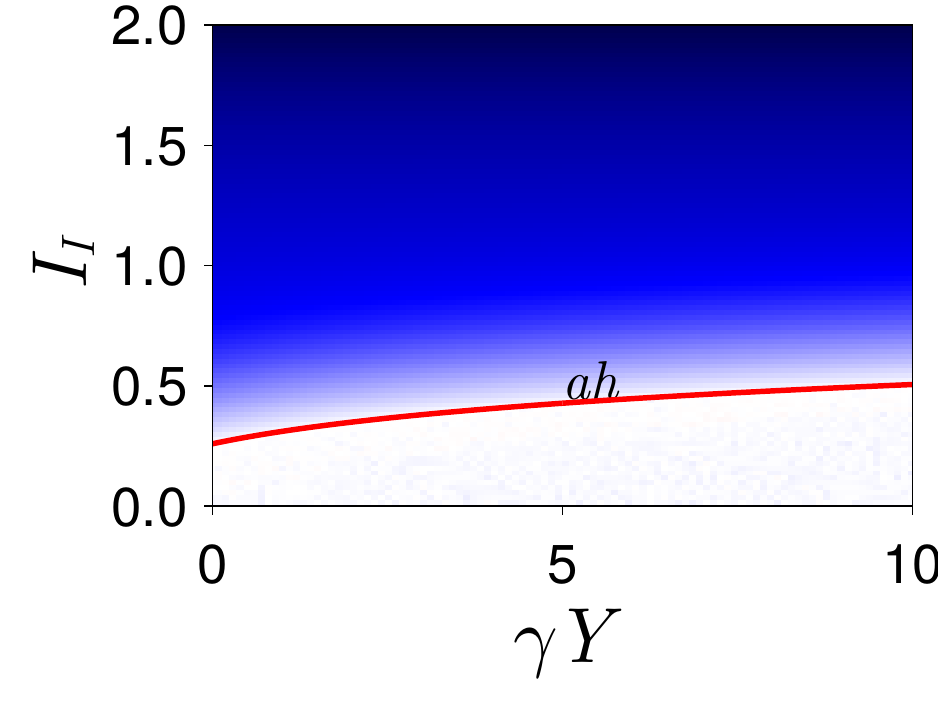}\\
    (c)\includegraphics[width = 0.45\linewidth]{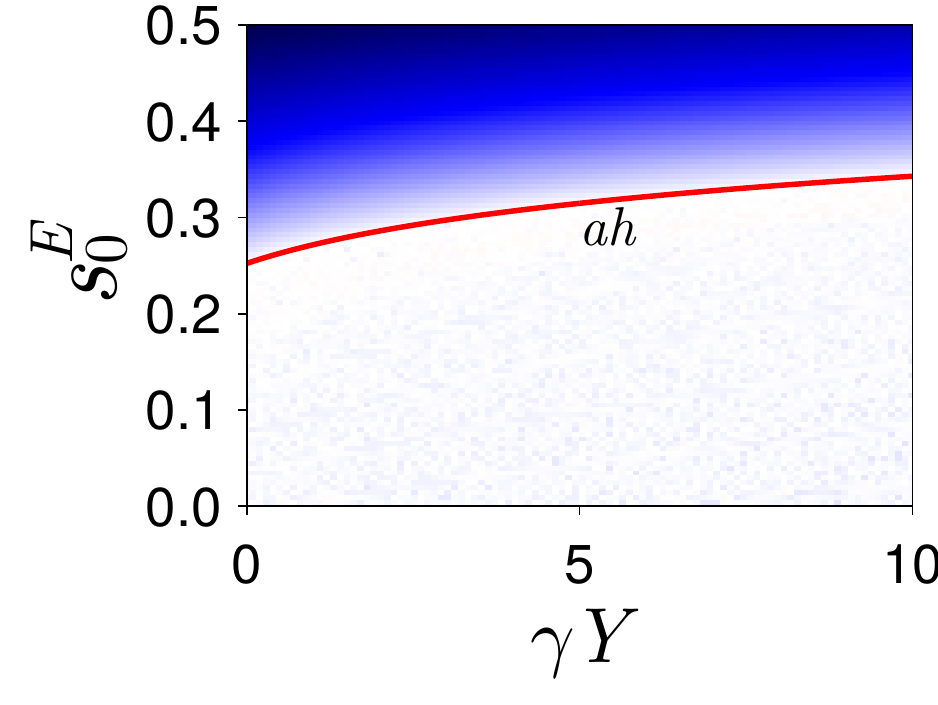}
    (d)\includegraphics[width = 0.45\linewidth]{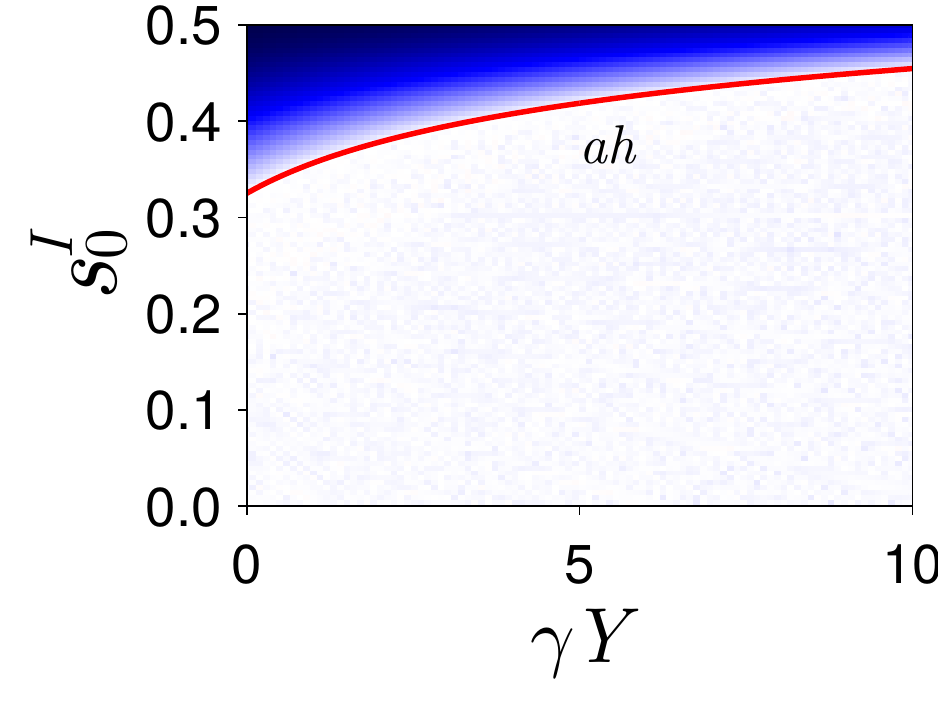}\\
	  \caption{Charts of MLE $\lambda$ on different parameter planes: (a) $(I_E,\gamma_Y)$, (b) $(I_I,\gamma_Y)$, (c) $(s^E_0,\gamma_Y)$, (d) $(s^I_0, \gamma_Y)$. Here blue regions corresponds to stable equilibrium (a "quiescence" state, $\lambda<0$), white regions -- to  stable limit cycle (a periodic oscillations, $\lambda=0$ and other Lyapunov exponents are negative). Here $ah$ label marks Andronov-Hopf bifurcation, $f_1$ and $f_2$ -- fold bifurcations. See detailed description in the text.}
 \label{fig:LLE_maps}
\end{center}
\end{figure}

To investigate the impact of astrocytes on gamma oscillations arising from the interaction of excitatory and inhibitory neuron populations on the example of system \eqref{eq:main}, we conducted a two-parameter bifurcation analysis of system \eqref{eq:main}. 

First of all, we constructed the chart of maximal Lyapunov exponent (MLE) $\lambda$ on the planes of control parameters $(I_E,\gamma_Y)$,  $(I_I,\gamma_Y)$, $(s^E_0,\gamma_Y)$ and $(s^I_0, \gamma_Y)$, see Fig. \ref{fig:LLE_maps}. 
This chart consists of different colored regions whose colors correspond to different regimes of this system. Regions with shades of blue correspond to those parameter values at which system \eqref{eq:main} has a stable equilibrium (a "quiescence" state, $\lambda<0$). White regions  correspond to those parameter values at which system \eqref{eq:main} has a stable limit cycle (a periodic oscillations, $\lambda=0$ and other Lyapunov exponents are negative). The red colored regions correspond to those parameter values at which system \eqref{eq:main} has chaotic dynamics (strange attractor, $\lambda>0$). 

The charts were computed using the following procedure. 
Using a limit cycle solution as a starting point, we perform attractor continuation by using a point from the attractor at the neighboring parameter values. 
Using this point as a starting point for a trajectory, we cut the transient (1000 units of time) and compute a Lyapunov spectrum during 5000 time units.
For computing Lyapunov spectrum we have used a numerical implementation from DynamicalSystems.jl \cite{Datseris2018} combined with 9th order Verner method from DifferentialEquations.jl \cite{rackauckas2017differentialequations} (both are packages of Julia programming language).

However, as shown on Fig.~\ref{fig:LLE_maps} this rough exploration of parameter space has found only quiescent or periodic dynamics despite the system's high dimension and usual abundance of chaotic or quasi-periodic dynamics in such systems.

The observed transition from periodic dynamics to stable equilibrium at Fig.\ref{fig:LLE_maps} suggests the involvement of equilibria in system's bifurcations. 
For parameter values close to this transition a numerical method for finding equilibria has been used.
The problem of finding equilibria was posed as finding a solution to the system of nonlinear equations \eqref{eq:TruncEquilibria}, which was solved by using a version of Newton method from IntervalRootFinding.jl package adapted to interval arithmetics \cite{IntArithmJl}.
These equilibria were then continued with respect to system's parameters using BifurcationKit.jl package \cite{veltz} and MATCONT software \cite{matcont}.

Fig.~\ref{fig:LLE_maps} shows the found bifurcation curves superimposed with MLE chart.
Here curves $ah$, $f_1$ and $f_2$ correspond to bifurcations of equilibria. The red curve $ah$ corresponds to an Andronov-Hopf bifurcation of the equilibrium state, which gives rise to different types of regular rhythmic activity. Curves $f_1$ and $f_2$ relate to the fold bifurcations.  Let us explain in detail bifurcation transitions on the example of Fig. \ref{fig:LLE_maps}(a). In order to do this let us study the route on this MLE chart, corresponding to some fixed value of $\gamma_Y$ and increasing $I_E$. For parameter values from blue region from the bottom of Fig. \ref{fig:LLE_maps}(a), only stable equilibrium $O_1$ exist. During the supercritical Andronov-Hopf bifurcation $ah$, it becomes a saddle one and a stable limit cycle $C_1$ is born in its vicinity. Further during fold bifurcation $f_1$ a stable equilibrium $O_2$ is born  together with corresponding saddle equilibria $O_3$. During fold bifurcation $f_2$ two saddle equilibria $O_1$ and $O_3$ merge and disappear. Note that curves $f_1$ and $f_2$ intersect at cusp point, while curves $ah$ and $f_1$ intersect at Bogdanov-Takens point, both of which lie outside the range of permissible parameter values. The found bifurcation curves show no further signs of complex dynamics in this system: both continuation of equilibria and limit cycles (not shown in these figures) do not demonstrate transition to either quasiperiodic or chaotic dynamics.

It is worth to note that for parameter values from the region between $f_1$ and $f_2$ curves a phenomenon of multistability is observed. Here a stable limit cycle $C_1$ coexists with the stable equilibrium $O_2$. 
Since system \eqref{eq:main} exhibits multistability, we also employed a brute force search in its phase space for other possible attractors that could have been overlooked during previous attractor continuations (and that might potentially demonstrate more complex dynamic behavior).
For that we took parameter values from equally spaced grids on four lines: $I_E = 0.9$, $I_I = 0.2$, $s_0^E = 0.1$ and $s_0^I = 0.2$, varying $\gamma_Y$ from $0$ to $100$ with a step equal to $0.1$ (other parameters' values are fixed and are the same as in the end of Section~\ref{sec:model}). Such lines completely lie in the regions on Fig.~\ref{fig:LLE_maps} where only an attracting limit cycle is present. 
Using the observations from Subsection~\ref{subsec:eq_states}, for each of these parameter values we generated $250$ random initial conditions in the hyper rectangle $\mathcal{H}$ using them as a starting point for trajectories. 
As before, first 1000 time units of a trajectory are considered as a transient, and the whole Lyapunov spectrum is computed during the next 20000 time units. 
The computed spectrum was then used as a way to distinguish potentially chaotic attractors from regular regimes with periodic behaviour.
However, during this brute force search no chaotic attractors were found: the maximal Lyapunov exponent in all cases was less than $10^{-4}$ or very close to zero, while the second largest was less than zero, which corresponds to an attracting limit cycle.

\subsection{Different types of rhythmic activity}
\label{subsec:rhytms}

In the paper \cite{keeley2019firing} devoted to rate model with synaptic dynamics without astrocytic influence it was shown that by introducing separate rate and synaptic timescales to the Wilson-Cowan framework, one can alter the amplitude profiles and oscillatory behavior of the gamma into weakly and strongly modulated regimes. Now let us analyze how astrocytes influence gamma rhythm that emerge in coupled neural populations described by Eqs. \eqref{eq:main}.

In previous Subsection we studied the mechanism of emergence and destruction of synchronized population activity when changing control parameters in system \eqref{eq:main}. In particular, we showed that in system \eqref{eq:main} in wide range of $\gamma_Y$ values, synchronization and gamma oscillations emerge from the interplay between the excitatory and inhibitory populations in the presence of astrocytes. Now let us examine how population activity patterns change with varying levels of astrocytic influence on neurotransmitter release $\gamma_Y$. 

\begin{figure}
\begin{center}
    (a)\includegraphics[width = 0.8\linewidth]{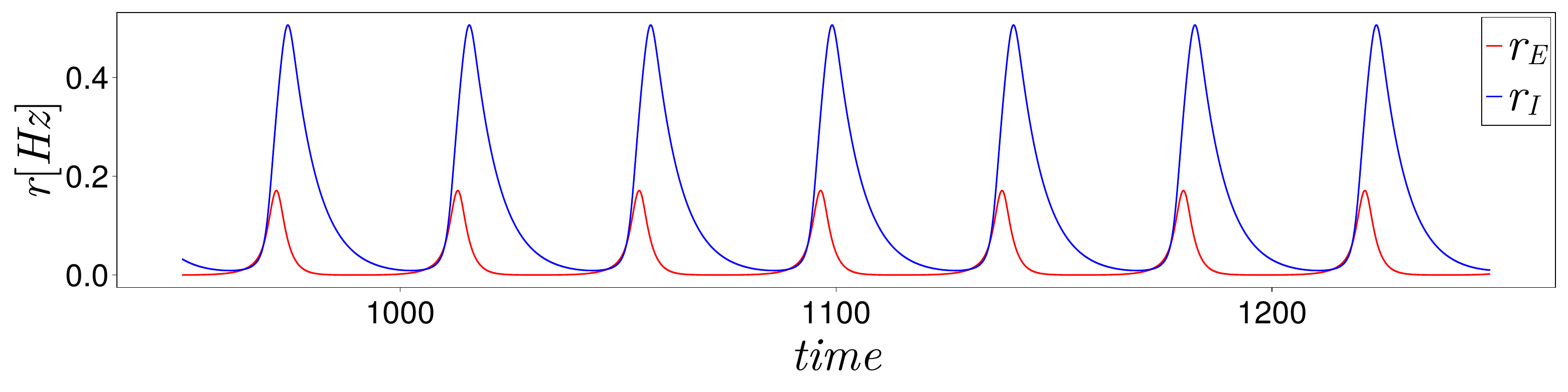}\\
    (b)\includegraphics[width = 0.8\linewidth]{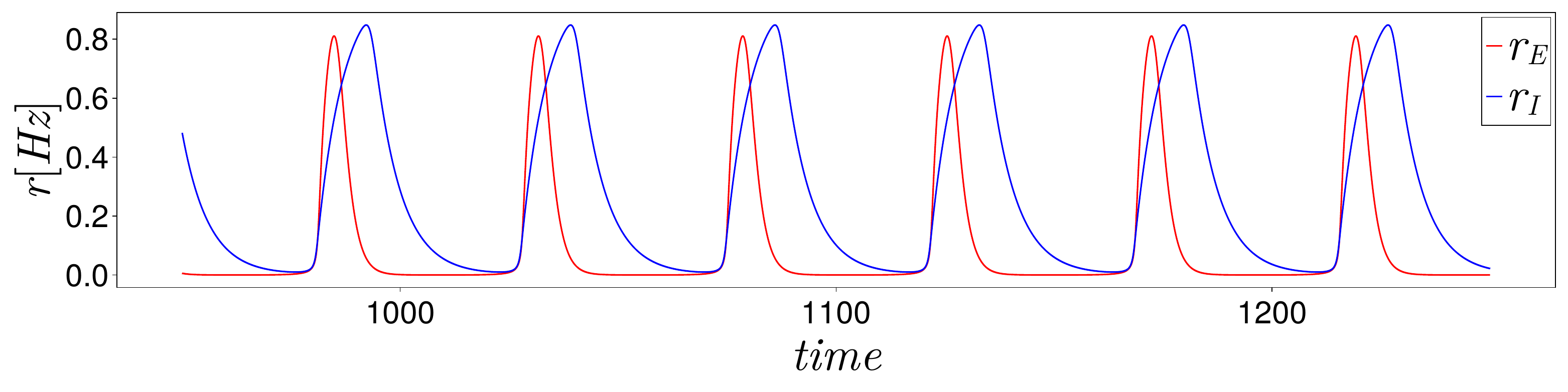}
    (c)\includegraphics[width = 0.8\linewidth]{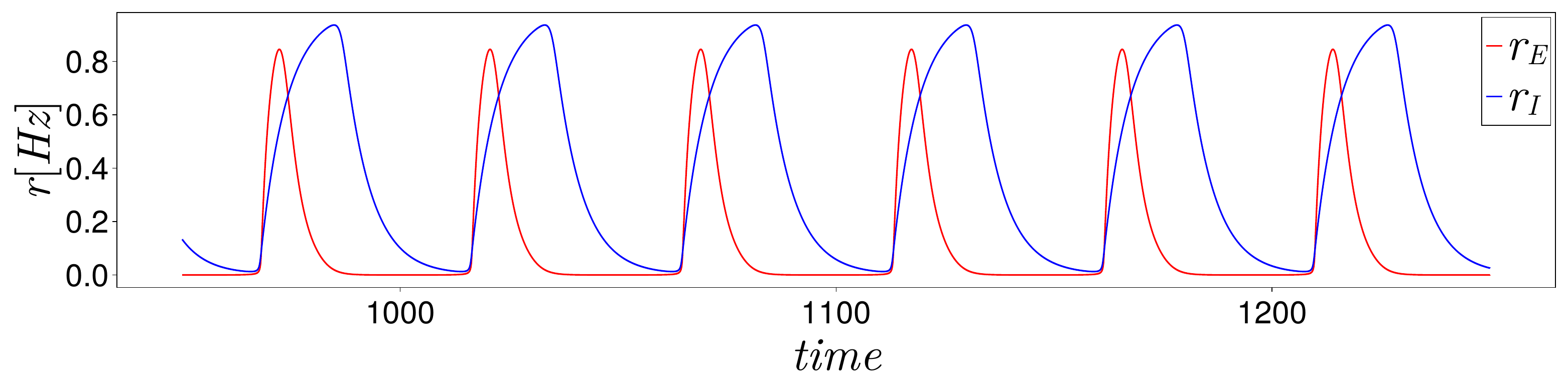}
    (d)\includegraphics[width = 0.8\linewidth]{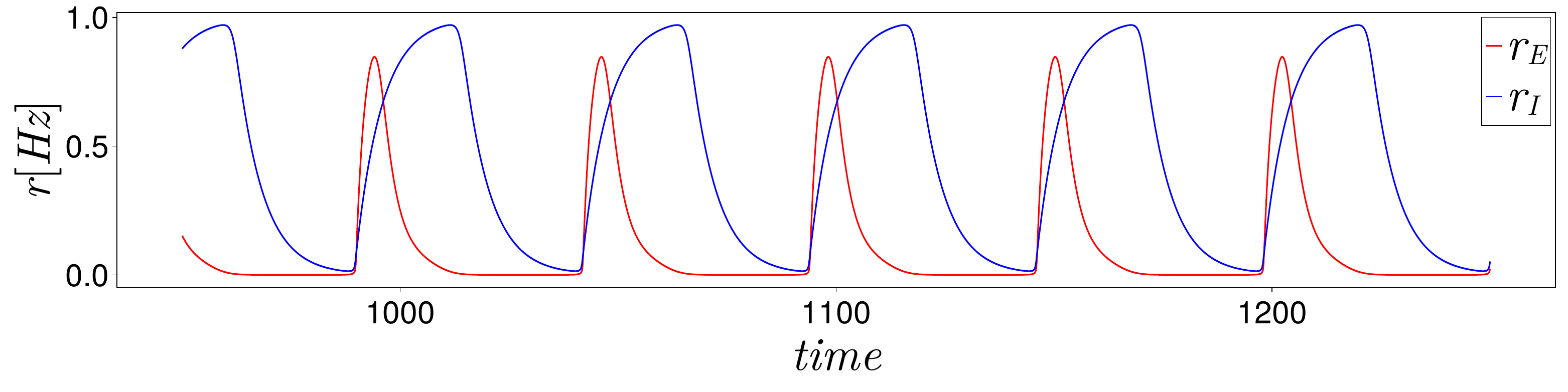}
	\caption{Time series of $r_E, r_I$. (a) $\gamma_Y = 2.5$, (b) $\gamma_Y = 10$, (c) $\gamma_Y = 50$, (d) $\gamma_Y = 100$ for fixed $s^I_0 = 0.325$. Red curve corresponds to the excitatory population of neurons, blue curve -- to the inhibitory population of neurons.} 
 \label{fig:diff_realisation}
\end{center}
\end{figure}

\begin{figure}
\begin{center}
     (a)\includegraphics[width = 0.8\linewidth]{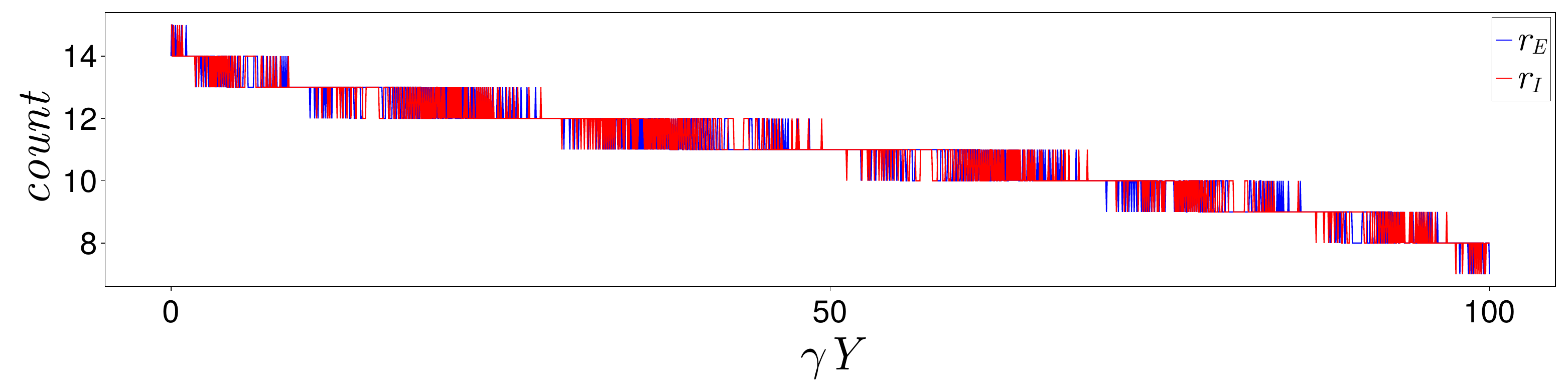}\\
     (b)\includegraphics[width = 0.8\linewidth]{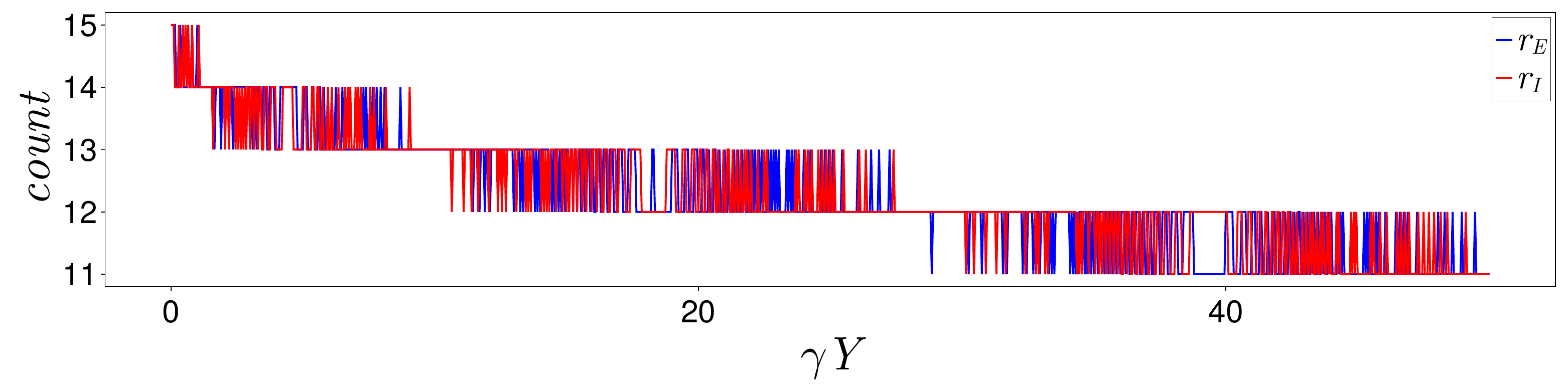}\\
     (c)\includegraphics[width = 0.8\linewidth]{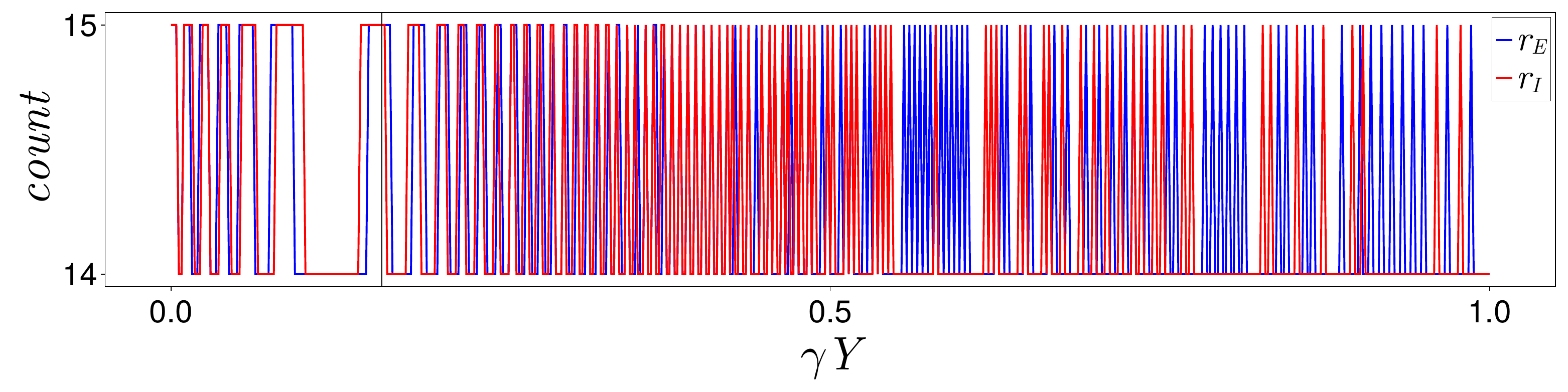}
\caption{Firing rates for $r_E, r_I$. (a) $\gamma_Y \in [0.0, 100.0]$. (b) $\gamma_Y \in [0.0, 50.0]$. (c) $\gamma_Y \in [0.0, 1.0]$. }
\label{fig:firingrates}
\end{center}
\end{figure}

In Fig. \ref{fig:diff_realisation} time series for rate variables $r_E$ and $r_I$ for different values of $\gamma_Y$ are presented. It can be clearly seen that astrocytic influence on neuronal activity intensifies with the increase in the value of $\gamma_Y$. For sufficient values of $\gamma_Y$, such as $\gamma_Y=100$ (Fig. \ref{fig:diff_realisation}(d)), the balance between excitation and inhibition is disrupted, which causes a decrease in oscillation frequency and amplitude, finally leading to oscillation death. Obtained results correspond to experimental data on the occurrence of excitotoxicity \cite{lebedeva2023effect}. Firing rates for $r_E, r_I$ and different $\gamma_Y$ are presented in Fig. \ref{fig:firingrates}.

From the point of view of biophysical modeling, the absence of chaotic patterns of activity in our model allows us to construct a correct description of the influence of healthy astrocytes on the dynamics of connected neuronal populations. Namely, the addition of healthy astrocytes to neuronal populations does not lead to the appearance of pathological (chaotic) dynamics; on the contrary, the astrocyte makes the dynamics of the system regular.

Note that observed results are determined by the existence of the feedback loop between neurons and glial cells and are independent of the complexity of the local dynamics of neurons and glial cells, specific characteristics of the neuron–glial interaction and a  particular architecture of the neural network.

\section{Conclusion}
\label{sec:conclusion}

In this study we have proposed a new mean-field model of brain rhythms controlled by glial cells. The novelty of our study is the extension of an enhanced Wilson-Cowan model that allows to take into account the impact of glial cells, particularly astrocytes, which can modulate the excitability of neurons by releasing chemical messengers called gliotransmitters. The proposed model provides a theoretical framework for understanding how glial cells can regulate neuronal activity and contribute to the generation of brain rhythms. To the best of our knowledge, these peculiarities have not been considered earlier.

The developed model allows to reproduce a variety of temporal patterns: from trivial ones, such as quiescence, to regular oscillatory activity. Mathematical images of these types of activity in the phase space of the proposed system were described. In particular, it was shown that regular oscillatory activity is connected to the appearance of periodic orbits in the phase space of the system under study. We have used bifurcation theory to obtain the mathematical description of transitions between the main types of population activity, caused by variations in the control parameters, first of all, by variations in parameter $\gamma_Y$ that characterize strength of astrocytic influence. It was shown that astrocytes can either enhance or inhibit the activity of neurons, depending on the specific type of glial cell and the conditions in which they are released.

The model allows us to demonstrate that the addition of a healthy astrocyte to the neural system makes its dynamics regular. In subsequent studies on the basis of the mean-field approach, we will describe how the dynamics of such a system will be affected by an astrocyte damaged, e.g., by a virus.

Overall, the mean-field model of brain rhythms controlled by glial cells provides a promising avenue for understanding the role of glial cells in the regulation of neuronal activity and brain rhythms. However, further research is needed to fully understand the complex interactions between neurons and glial cells in the brain, and to determine more precisely the specific mechanisms by which glial cells contribute to the generation and modulation of different brain rhythms. 

{\bf Acknowledgement.}

This work was supported by Russian Science Foundation, grant \# 19-72-10128.

\section*{Declaration of competing interest}

The authors declare that they have no known competing financial interests or personal relationships that could have appeared to
influence the work reported in this paper.

\section*{Data Availability Statement}

The data that support the findings of this study are available from the corresponding author upon reasonable request.

\bibliographystyle{epj.bst}
\bibliography{biblio}

\end{document}